# Energy Efficient Wireless Networking and Computing Infrastructure

Ahmed MATEEN[1]*          Sammar ABBAS[1]
[1]Department of Computer Science, University of Agriculture, Faisalabad, Pakistan

**Abstract**

Wireless networking allows users to access information and services regardless of location and physical infrastructure. It is a fast growing technology due to its availability of wireless devices, flexibility, ease of installation and configuration. With this rapid expansion of information and Communication Technology (ICT), the consumption of energy is also increasing. In the early age of wireless technology, computing infrastructure focused on everywhere access, capacity and speed of technology. But now computing infrastructure should be energy efficient because, in wireless networking, devices are mostly powered by a battery that is a limited source of energy and is a challenge for the researchers. In computing infrastructure energy saving and environmental protection has become a global demand. This paper proposed a computing infrastructure based on green computing for energy efficient wireless networking. Further, some challenges and techniques like power consumption in network architecture, algorithm efficiency, virtualization, and dynamic power saving will be discussed to make energy efficient computing infrastructure.

## INTRODUCTION

Two or more computers connected together for information sharing is called computer network. There are two main categories of computer networks namely wired and wireless network. Wired networks use wires as a communication medium while Wireless network allows users to access information and services regardless of wires, location and physical infrastructure using radio communication based on IEEE 802.11 group's standard. It is a fast growing technology due to ease of installation, configuration and availability of wireless devices[1].

In the early age of wireless technology, computing infrastructure focused on everywhere access, capacity and speed. But now, computing infrastructure should be energy efficient because in wireless networking devices are mostly powered by a battery that is a limited source of energy and is a challenge for the researchers. In computing infrastructure, energy saving and environmental protection has become a global demand because it is beneficial to all firms and individuals as well as to the environment. This paper proposed a computing infrastructure based on green computing for energy efficient wireless networking. Further some challenges and techniques like network architecture, algorithm efficiency, resource allocation, virtualization, power consumption and dynamic power saving aspects will be discussed to make energy efficient computing infrastructure. Wireless networks are classified into two categories namely structured (infrastructure mode) and structureless (ad-hoc mode) based on architecture.

### Infrastructure Mode

The first type structured network has base wireless stations and fixed Access Points (AP's). AP's usually connected through the wire with ISP (Internet Service Provider) to provide internet services. In this mode AP's are fixed and provide wireless signals to mobile nodes. Each AP has SSID (Service Set Identifier) and each node get authorization by that SSID to join the network. The nodes are free to move within the coverage area of AP's.

Three laptops are shown in Fig. 1 connected with the access point and the access point is connected with a modem through a wire that modem provide internet services. In this fig laptop (A) send information to AP and AP send that information to the laptop (B) and vice versa. Infrastructure mode wireless networks are centralized, scalable and secure.

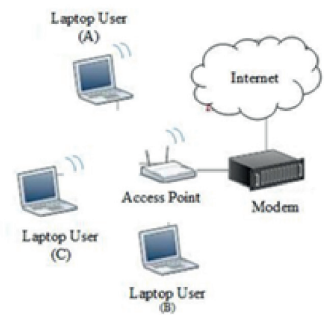

**Figure 1.** Infrastructure Mode of Wireless Network

### Ad-Hoc Mode

The second type wireless network is a decentralized wireless network that includes two or more wireless stations to communicate without any preexisting infrastructure, central control, and access points. Ad-hoc mode is self-configuring which can form dynamic topology[2]. In this mode, all nodes or devices act as a router by forwarding data for other nodes. It allows nodes to communicate even when they are not in range of each other directly. Packets are exchanged between these nodes by the intermediate node.

Figure 2 in Part (A) shows that cell phones are directly connected to each other without any AP and fixed infrastructure. Both cell phones are moving in different directions after some time as shown in Part (B). They are out of range from each other but they are still communicating. The laptop is working as an intermediate node between them and exchanging their messages to each other. Ad-hoc networks are suitable because of light configuration and fast implementation in emergency situations like natural disasters and military operations etc.[3].

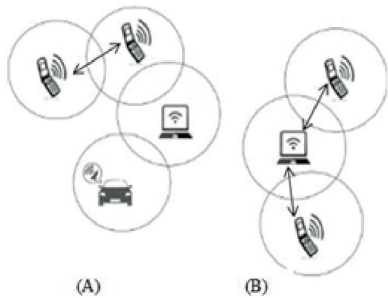

**Figure 2.** Ad-hoc Mode of Wireless Network

**Wireless Applications**

*Military Communication*

Ad-hoc networks are also used in military communications. In this modern age military tools, consists of some electronic and computer enabled system and devices to develop fast communication network between the soldiers, vehicles, and military headquarters.

*Emergency Operations*

The ad-hoc network also used in emergency operations where the communication system is damaged and quick implementation of a communication network is required. E.g. fire, flood, and earthquake situations. Information is conveyed from one member to another member of rescue team through a small handheld machines.

*Civil Applications*

Ad-hoc networks are also used in our civil life like at our home, education systems, conference rooms and meeting rooms to share information in short time.

**Components of Wireless Computing Infrastructure**

Wireless networking infrastructure has a large number of components or devices. It may include a personal computer, laptop, cell phones or any other handheld device to use wireless services and other required component is a wireless card which is Wi-Fi enabled. Router and switch are used to connect multiple computers with each other. Each wireless device uses an Ethernet port to connect with another system, computer, and device. Mostly new coming personal computers and laptop have integrated Ethernet adapters and simply require connection of telephone line, cable or DSL modem. In wireless network, a computer can be connected to a modem through a wireless switch or access point.

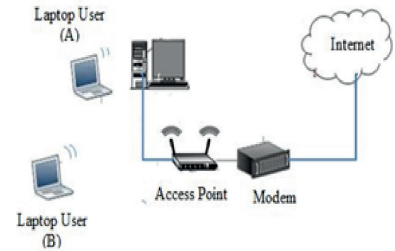

**Figure 3.** Wireless Network Infrastructure

**Green Computing**

Green is an English word and it describes a color having pleasant effect to the human eye. The term green computing is the use of computers and their resource with no impact on the environment. Today almost all fields like IT, medicine, transport and farming use tools which need lots of power and money for their active performance. People have excellent devices and tools to achieve their projects. The objective of green processing is to lessen the use of risky resources and energy consumption in real life. It promotes the recyclability or biodegradability of manufacturer waste. The Environmental Protection Agency (EPA) launched the Energy Star program in 1992 and started work on the green computing. After that many manufacturers are designing energy efficient device to minimize the environmental impact[4].

In Computing Infrastructure, energy saving and environmental protection has become a global demand because it is beneficial to all firms and individuals as well as to the environment. This paper proposed energy efficient wireless networking and computing infrastructure based on green computing concepts.

## RELATED WORK

Green computing is a hot topic for research now a days and many researchers use their efforts to make effective use of computer power and are working on the right way to use electronic devices. The one most device being used is a personal computer and they considered it for their research. They focus on energy consumption of different components of personal computer like memory, processor and monitor[5].

In the high progress of wireless networking, there are many number of wireless networking techniques and topologies for the communication. They considered 802.11 ad-hoc modes for the direct communication and stated as poor communication services. They proposed MA-Fi multi-hop mobile networking with the IEEE 802.11 infrastructure mode using virtualization in network topology MA-Fi is applicable and reduces energy consumption up to 60 to 70 %[6].

Macrocell is the base station mostly used in wireless networks. However, interest is increasing in trying different approaches for different scenarios. Microcells are small areas used to fill coverage gaps. Moreover, the service providers are focusing on home base stations that can be used to transfer subscribers from fixed line to wireless equipment. For providing services in planes or on ships, Femto-cells are being considered. Cost effectiveness in terms of power and traffic capacity becomes a requirement. Their paper identified the resources necessary to reach an overall RF system idea that is controlled and is compliant with other market requirements[7].

There is a big need of environmental friendly and energy saving computing devices. The green computing represents responsible way to reduce power consumption and power bill. It saves the country resources as a whole. Green technologies are available for a single user or an organization. Both can take advantages of green computing and can play role to save environment[8].

## MATERIALS AND METHODS

As discussed in introduction that the computing infrastructure include number of devices like computer, laptops, servers, cell phones and other handheld devices used with the help of operating systems and applications. Between two or more devices the networking concept/infrastructure is used. Networking infrastructure basically consists of a sender, receiver and transmission medium. The wired networks use wires as medium while wireless networks use radio waves as medium as shown in fig 4.

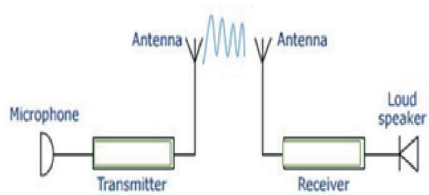

**Figure 4.** Wireless Network Module

With the above given basic modules computing infrastructure has too many supporting devices. The proposed study focused on the user devices (sender and receiver) the medium of wireless computing infrastructure with respect to energy consumption and suggested different approaches to minimize the energy consumption as given follows.

The maximum amount of energy is unexpectedly used in the idle state of nodes. The second main energy consumption is due to obstacle and interaction of other materials. The third cost is due to overhearing. Finally, the power utilization in the Transmit and Receive states are very low. With all these, a lot of work and research is needed to minimize the energy lost at different steps of the network to increase the network lifetime.

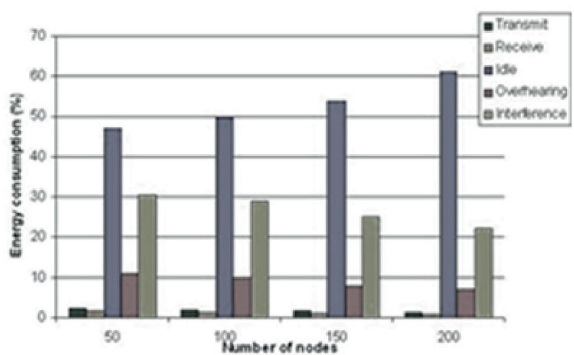

**Figure 5.** Energy Consumption without Sleep State

In the above-given scenario, as shown in figure 6, both users are moving in the simulation. Traffic is created among users through the access point. The access point acts as transmitter and receiver.

The power monitoring is done by the simulator NS-3. The figure shows the graph of remaining energy of access point. As the time passes the remaining energy graphs goes down. The energy consumption of nodes is shown as transmitter Tx and receiver Rx. Fig. 7. and Fig. 8. shows Energy consumption of both nodes on Y-axis is increasing with passage of Time at X-axis. In fig. 7. the Node 2 consuming more energy than Node 1 because of the distance from the AP. While in fig. 8. both nodes using same energy while receiving data packets from AP. Here power is consumed by AP for sending data.

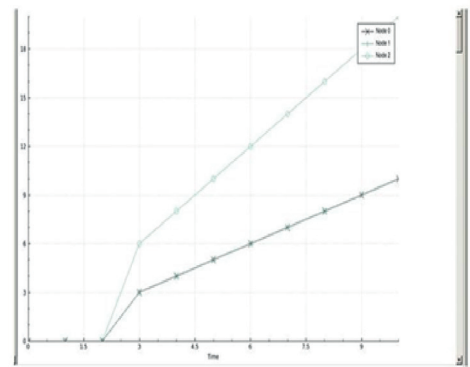

**Figure 7.** Energy Consumption of Nodes (Tx)

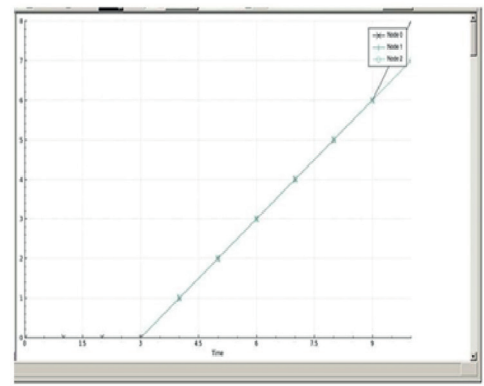

**Figure 8.** Energy Consumption of AP and Nodes (Rx)

The computing infrastructure includes multiple devices to provide services. All of them need the energy to perform their functionality. Due to increase in users and a large number of its applications, energy consumption is also increasing. The energy consumption cause high electrical bill and environmental effects so there is a need to control the energy crisis in our country Computing infrastructure can be energy efficient by following techniques

### Product Manufacturing

The manufacturing process of computer and other networking devices is mostly based on natural resources used in the life cycle. Consequently, the extension in lifetime of any device supports the green computing. Dell, HP and Apple are the main computer devices producers are producing energy efficient with Energy Star. The Energy-Star, hardware products are energy efficient standards. These days, most PCs are manufactured with a sleep approach that permits them to shut down when not being used and, in this way it saves energy. Manufacturing another PC through scratch has a much environmental effect than a new module.

**Product Designing**

In the product designing, energy efficient devices are designed. In the early age of computers the size of computer hardware was too much large and the consumption of energy was also very high. Then by improving the designing techniques the size of computers become small and the along with it, the consumption of energy also decreased. But it's not enough it should be more minimized. A lot of work has been done to make the good user-friendly design of devices and software interface. Integrated circuits helped so much in this aspect. The size of device and energy consumption are reduced at a high level So that we have a handheld device and is continuously decreasing.

**Product Usage**

In wireless network accesses points, base stations and in systems processor are the main consumer of energy, and, therefore, we first create a load on the processor. It is found that 70% processor used in idle state and 30 % inactive or working. The usage of the device is most critical in energy efficiency. When you are not using your system, shut down the system. It's better than you leave it or it goes idle state or in sleep mode. Such as Cathode Ray Tube (CRT) is the first type of monitors. They have well-looking angles. But they consume high power. On another hand, Liquid Crystal Display (LCD) monitors are in the market now for low power utilization. Such energy efficient tools can be very helpful to make energy efficient network.

**Resource Adjustment Strategies**

The improper placement of resources causes wastage of energy. While adjusting the hardware like multiple antennas and access points the coverage area should be checked e.g. it is covering the required area or not and how much area is overlapping with the adjacent antenna. Use Omni-directional or Uni-directional antenna according to the scenario. The capacity of the antenna should also be kept in mind; it should be according to traffic or number of users.

**Algorithm Efficiency**

The algorithms efficiency has an effect for any given computing job and there are different ways to write algorithms. As its applications increased the amount of energy consumption also increased. Algorithm efficiency can play a great role in this regard. Resource allocation algorithms can also be used to send data packets and data hubs where energy price is less. There are two main routing protocols namely Proactive and Reactive. Proactive routing protocols try to keep latest routing data, they answer to each variation in the network to keep network view[9].

Certain proactive routing protocols are Destination Sequenced Vector (DSDV), Wireless Routing Protocol (WRP). Reactive routing protocols are formed on request or when compulsory. When a source needs to send data to a destination, it appeals the route to find the path. Once a path has been discovered, it is upheld until there is no more need of that path. Certain proactive routing protocols are Ad-hoc On-Demand Distance Vector (AODV), Dynamic Source Routing (DSR) and Temporally Ordered Routing algorithm (TORA). In this paper comparison had been made between different routing algorithms means of simulation to check the total system energy consumption. Simulation performed on 50 Nodes with 512 bytes packet size for EAODV, AODV, DSDV, DSR, and TORA routing protocol. Results are shown in fig. 9 and fig. 10 with energy consumption on Y-axis, data rate and pause time on X-axis. TORA consume less energy during packet sending than other routing protocols as data rate of packet sending increases the consumption of energy decreases as shown in fig. 9. TORA consumes less energy in idle state also than other routing protocols as shown in fig. 10.

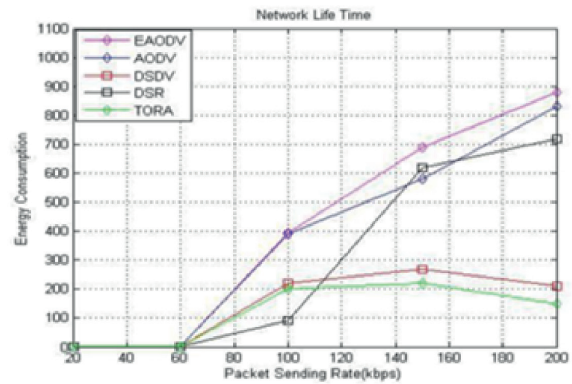

**Figure 9:** Energy Consumption during packet sending

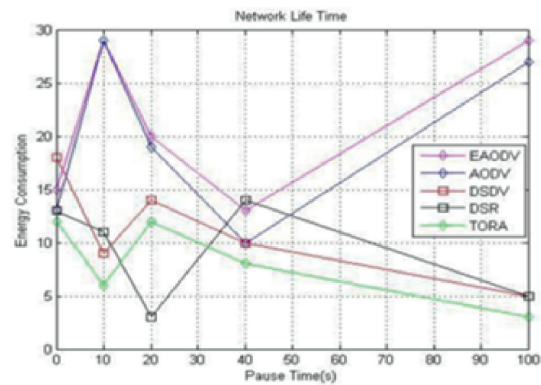

**Figure 10:** Energy Consumption during Pause/Idle state

**Virtualization**

With virtualization, a system administrator can join many physical systems into virtual machines on a one powerful system, so reducing the usage of hardware at large amount. It has an impact on energy consumption. Different profitable firms suggest different software packages to enable virtual computing. Cloud Computing is more popular computing nowadays. Cloud infrastructure service like data centers has a major role in providing these services such as Amazon, Google, Microsoft etc. Datacenter provides a central role to Software services providers for computing the reliable services and provides these services to end user by using the internet. The service which has been provided to end user is called "Software as a Service (SaaS)", and the service provides to software services provider by datacenter is called "Infrastructure as a Service (IaaS). The energy consumed by 10 virtual servers is equal to 1 physical server It reduces the hardware manufacturing by running multiple operating systems on a single hardware. It reduces implementation and maintenance cost also (Microsoft corporation, 2008).

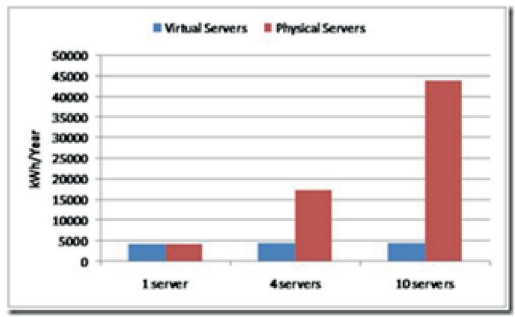

**Figure 10:** Virtual and Physical Server

## CONCLUSION

Wireless computing infrastructure consists of a large number of devices. This research concludes that which devices consume more power, how energy consumption can be reduced and which devices should be used. We found that in the wireless network the most energy consuming devices are base station, access points, and processor in an idle state. Similarly, many parts of energy consumed by Antennas for the propagation of signals are wasted due to directional issues or improper adjustments of antennas. We analyze that devices use maximum power in the idle state as shown in fig. 5 and we can reduce it by using TORA routing protocol. All protocols perform differently in different scenarios while with mobility or load increasing TORA energy consumption is low than others as shown in fig. 9 and 10. Correct selection of devices, adjustment, installation and correct usage of hardware and software make the computing infrastructure energy efficient and eco-friendly. Do not leave your computer ON at night and on weekends. Use virtualization in an enterprise or in the organization to reduce hardware, save your place, cost, and energy as shown in figure 10. This study suggests use of Energy Star products.

## REFERENCES

[1] Chansu Y., Ben L., and Yong H., 2003, Energy efficient routing protocols for mobile ad hoc networks. Wireless Communication Mobile Computing, 3, 959–973

[2] Alotaibi E., and Biswanath M., 2012, Survey paper A survey on routing algorithms for wireless Ad-Hoc and mesh networks. Computer Networks, 56(2): 940-965.

[3] Tuteja A., Ranjeesh G., and Sunil T., 2010, Comparative Performance Analysis of DSDV, AODV and DSR Routing Protocols in MANET using NS2. International Conference on Advances in Computer Engineering, 330-333.

[4] Shinde S., Nalawade S., and Nalawade A., 2013, Green Computing: Go Green and Save Energy. International Journal of Advanced Research in Computer Science and Software Engineering, 3(7):1033-1037.

[5] Mala A., Umarani C., and Ganesan L., 2013, Green Computing: Issues on the Monitor of Personal Computers. International journal Of Engineering And Science, 3(2): 31-36.

[6] Wirtz H., Georg K., Johannes L., Robert B., and Klaus W., 2014, High-performance, Energy-efficient Mobile Wireless Networking in 802.11 Infrastructure Mode. IEEE 11th International Conference on Mobile Ad Hoc and Sensor Systems, 291-299.

[7] Jeffrey S., Joens Y., Raluca S., 2014, Using Big Data to Improve Customer Experience and Business Performance. Bell Labs Technical Journal, 18(4): 3-17.

[8] Kiruthiga P., and kumar V., 2014, Green Computing – An Ecofriendly Approach for Energy Efficiency and Minimizing E-Waste, International Journal of Advanced Research in Computer and Communication Engineering, 3(4): pp. 6318-6321.

[9] Mohamad N., Usop A., Ahmad F., and Amri A., 2009, Performance Evaluation of AODV, DSDV & DSR Routing Protocol in Grid Environment, IJCSNS International Journal of Computer Science and Network Security, 9(7).